\documentclass[cup5b]{caps}
\usepackage{graphicx}
\usepackage{amssymb}
\usepackage{ociwsymp3e}  

\HeadText{S. C. Ellis and L. R. Jones}  

\begin{document}

\pagenumbering{arabic}

\author[]{S. C. Ellis and L. R. Jones,\\The University of Birmingham, UK}

%
%

\chapter{The \emph{K} Band Luminosity Function \\ of High Redshift Clusters}

\begin{abstract}

\emph{K} band observations of the galaxy populations of three high redshift ($z=0.8$--$1.0$), X-ray selected, massive clusters are presented.  The observations reach a depth of $K \simeq 21.5$, corresponding to $K^{*}+3.5$ mag.  The evolution of the galaxy properties are discussed in terms of their \emph{K} band luminosity functions and the \emph{K} band Hubble diagram of brightest cluster galaxies.

The bulk of the galaxies, as characterised by the parameter $K^{*}$ from the Schechter (1976) function, are found to be consistent with passive evolution with a redshift of formation of $z_{f}\approx 1.5$--2.  This is in agreement with observations of other high redshift clusters, but in disagreement with field galaxies at similar redshifts.  The shape of the luminosity function at high redshift, after correcting for passive evolution, is not significantly different from that of the Coma cluster, again consistent with passive evolution.

\end{abstract}

\section{Introduction}

We present a study of three of the most massive ($\sim 10^{15}$M$\odot$), high redshift clusters known (Maughan et al. 2003a, Maughan et al. 2003b).  They are thus ideal probes of galaxy evolution.  In a hierarchical model galaxies are predicted to first form in regions with the highest overdensities which merge
over time with other systems to become massive clusters.  Thus such massive clusters at high redshift are relatively rare and we have an unusual opportunity to
study the galaxy populations of rich, distant clusters and compare results with
local rich clusters such as Coma.  The high redshift of the clusters should make any evolution in the galaxy populations easier to observe.

The evolution is probed by means of their $K$ band galaxy luminosity functions.  The $K$ band magnitude of a galaxy is a good indicator of the stellar mass of the galaxy being relatively insensitive to star formation, and furthermore $k$-corrections are small in this band.  Thus the evolution of the $K$ band luminosity function, parametrized by $K^{*}$ of the Schechter (1976) function, traces the epoch of assembly of the galaxies.  In the monolithic collapse picture of structure formation all galaxies were formed at high redshift and have evolved only passively since, with galaxies at high redshift being intrinsically brighter than their present day counterparts due to their younger stellar populations.  In the hierarchical formation picture the number of bright galaxies will grow with time through a series of mergers and thus the shape of the luminosity function will be altered with the passing of time.

The X-ray data for the clusters show that two of the clusters appear relaxed (ClJ1226, $z=0.89$ and ClJ1415, $z=1.03$) with relaxed X-ray contours (Maughan et al. 2003b, Ebeling et al. 2001) and one of them (ClJ0152, $z=0.83$) appears to be in a state of merging (Maughan et al. 2003a, Ebeling et al. 2000), thus we also have a small
selection of different environments.  The cluster X-ray properties are consistent with little or no evolution when compared to local clusters (Jones et al. 2003, these proceedings).
All three clusters were discovered in the Wide Angle {\sc Rosat} Pointed Survey
(WARPS, Scharf et al. 1997).

\section{Results}

\begin{figure}
\begin{minipage}[c]{0.5\textwidth}
    \centering \includegraphics[scale=0.23,angle=270]
    {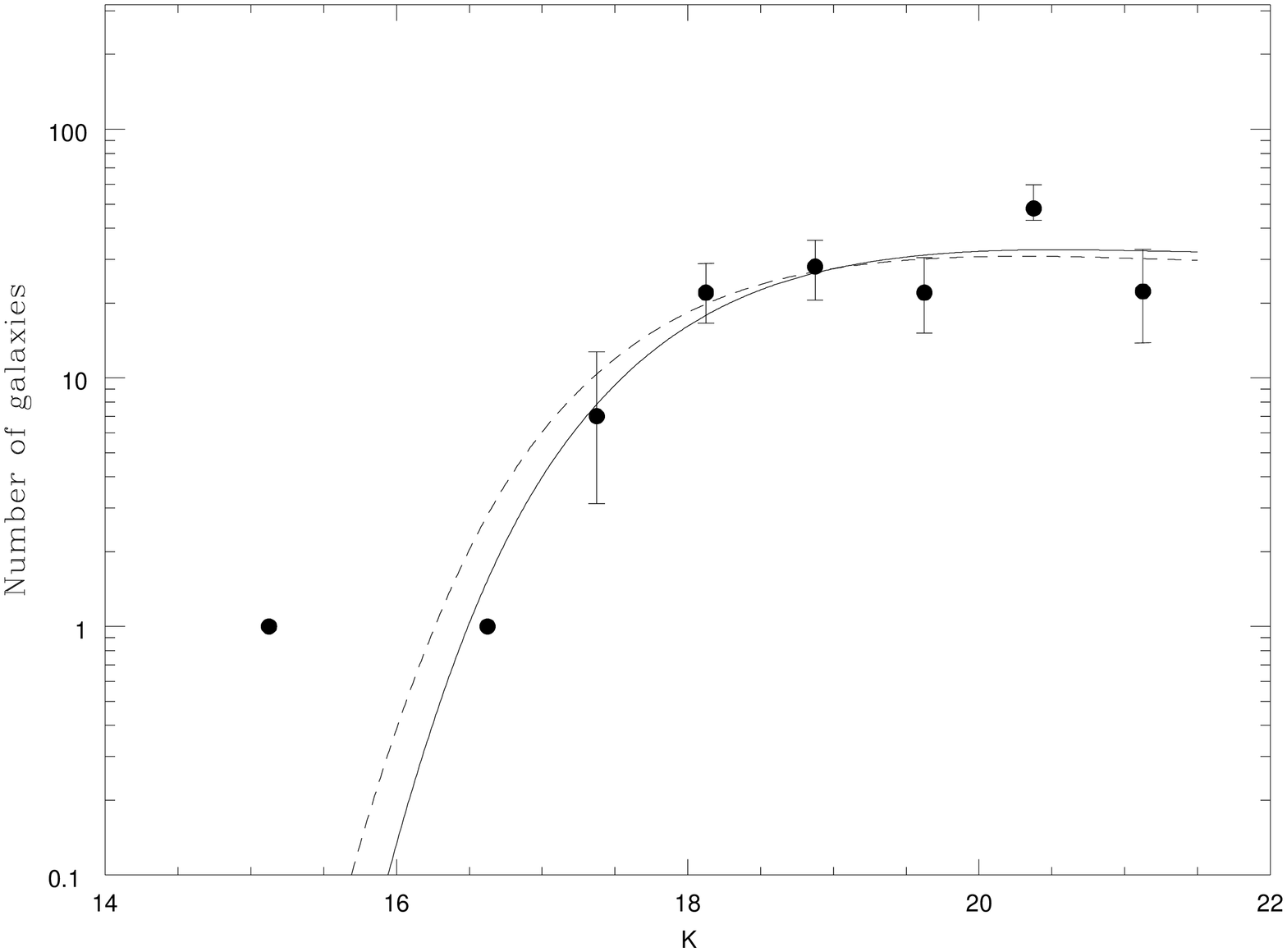}
  \end{minipage}%
  \begin{minipage}[c]{0.5\textwidth}
    \centering \includegraphics[scale=0.23,angle=270]
    {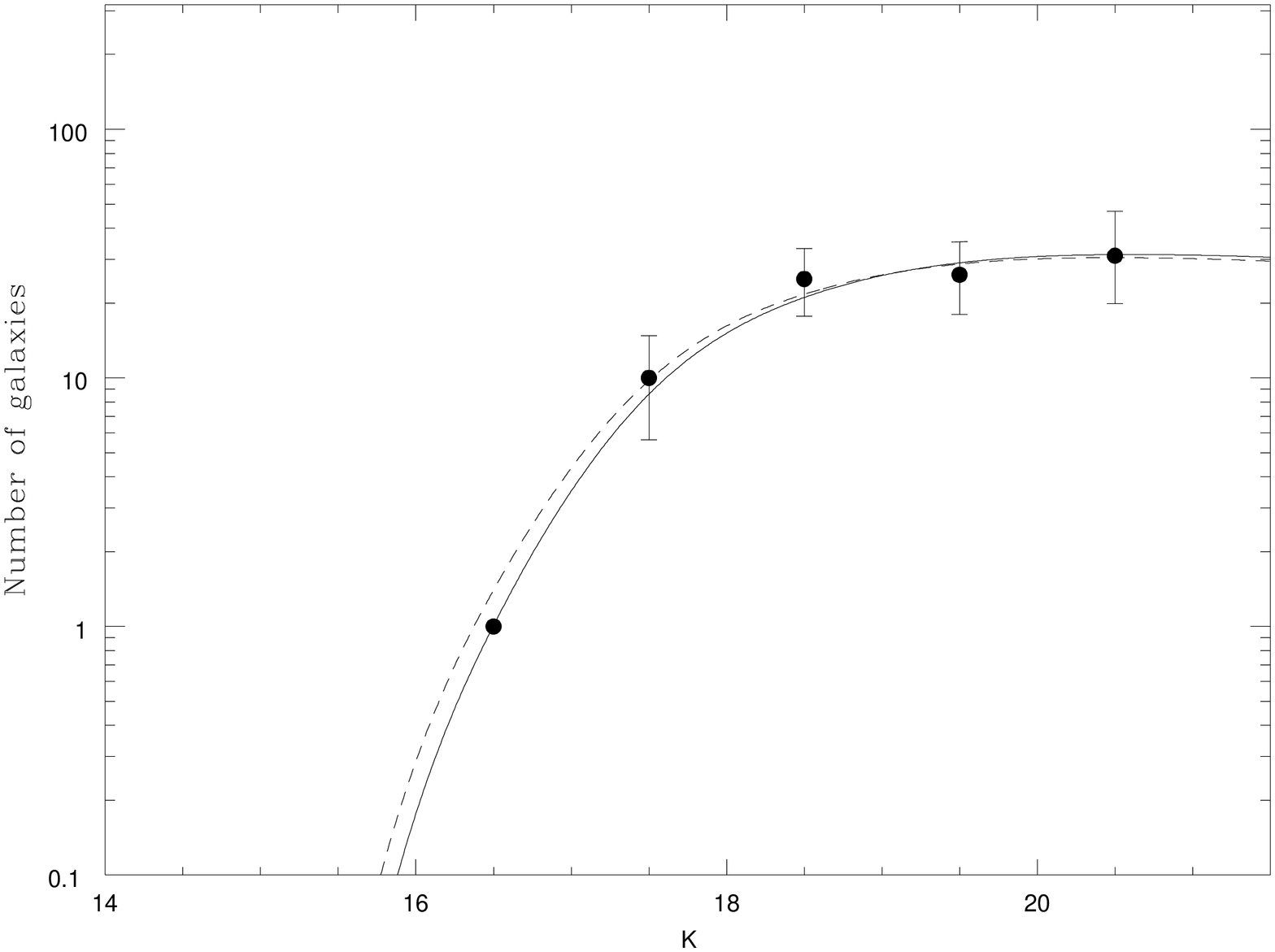}
  \end{minipage}
  \begin{minipage}[c]{0.5\textwidth}
    \centering \includegraphics[scale=0.23,angle=270]
    {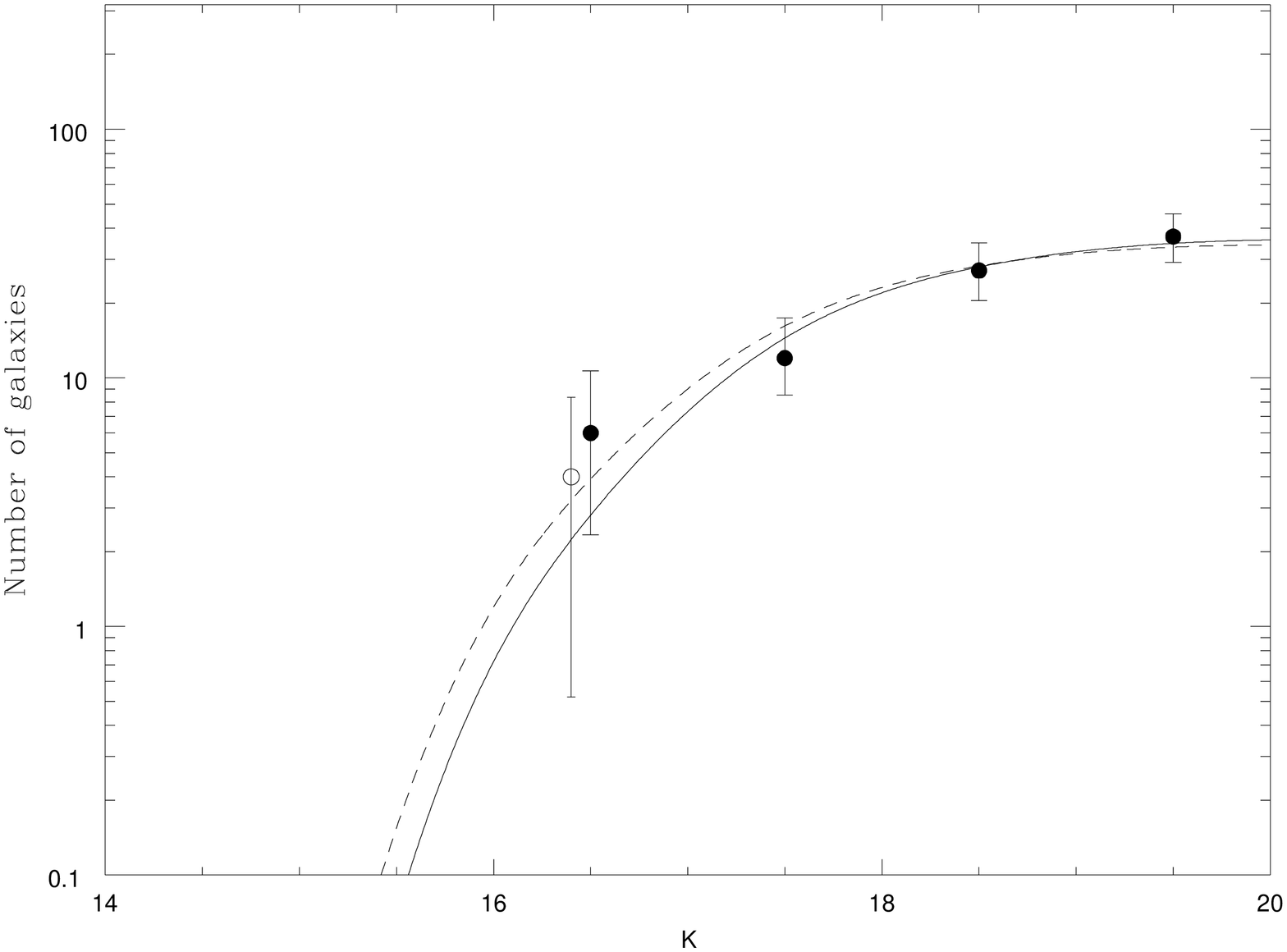}
  \end{minipage}%
  \begin{minipage}[c]{0.5\textwidth}
    \centering \includegraphics[scale=0.23,angle=270]
    {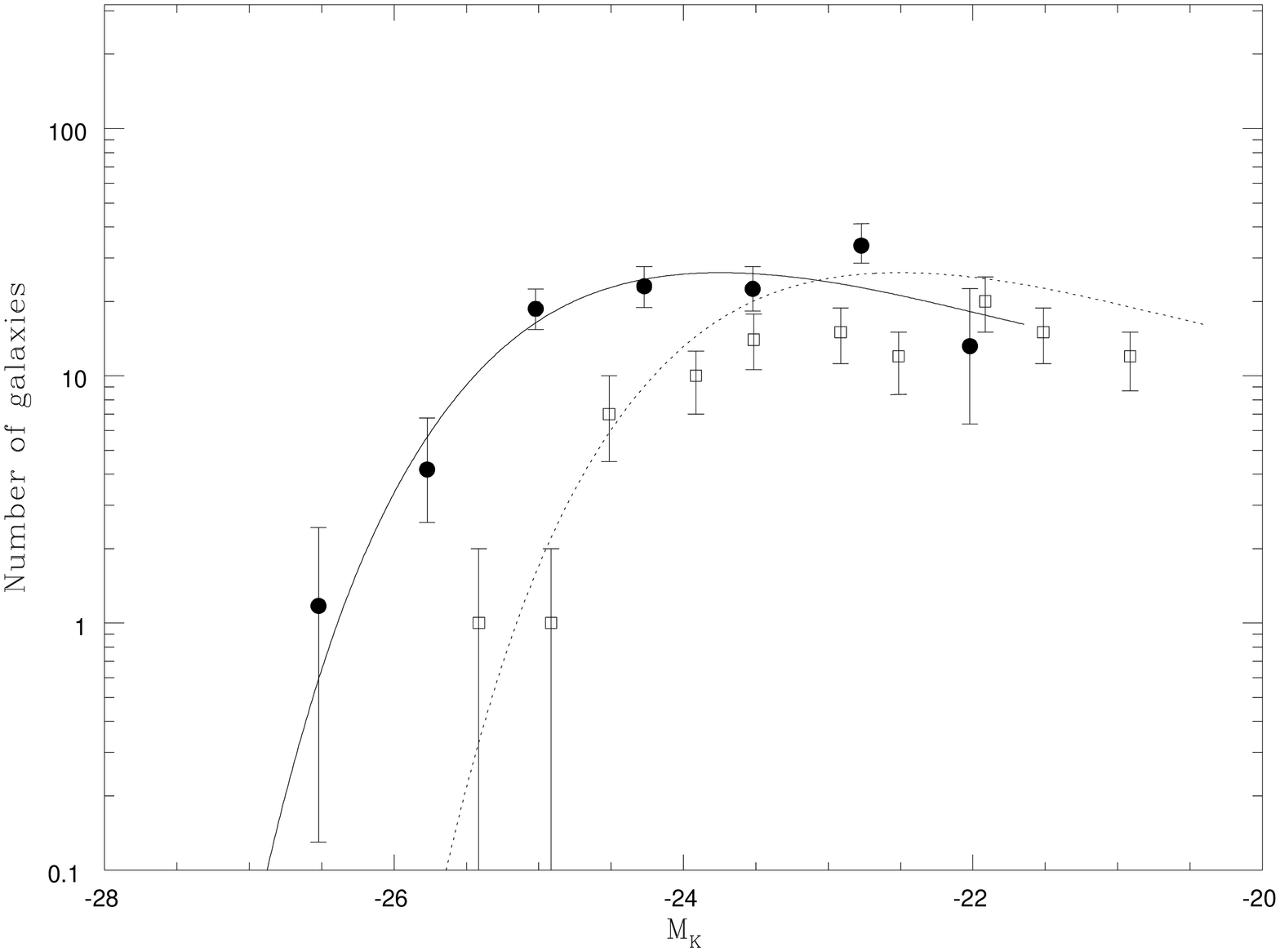}
  \end{minipage}
  \caption[Caption for TOC]
    {K Band luminosity function for ClJ1226, ClJ1415, ClJ0152 and the combined luminosity function of all three clusters compared to that of Coma.}
  \label{fig:klfs}

\end{figure}

The luminosity functions of all three clusters are shown in the first three panels of figure~\ref{fig:klfs}.  Also shown are the best fitting Schechter functions with $\alpha$ held constant at $\alpha=-0.9$.  Schechter functions were fit with the brightest cluster galaxies included and excluded.  The results for all clusters are given in tables~\ref{tab:schechter_fits} and \ref{tab:schechter_fits_bcg}.

The combined luminosity function, excluding BCGs, for our 3 high redshift clusters is shown in the bottom right panel of figure~\ref{fig:klfs}.  This was fit with $\alpha$ as a free parameter yielding a value of $\alpha=-0.54 \pm 0.3$.    Also shown, by the open squares, is the $K$ band luminosity function for the Coma cluster computed from the $H$ band luminosity function  of de Propris et al. (1998) using their given value of $H-K=0.22$.  The shapes of the luminosity functions are very similar and the difference in absolute magnitude can be accounted for by a passively evolving population with $z_{\rm{f}}=2$ as shown by the dashed line.

A Schechter function was also fit to the combined luminosity function including BCGs.  Fits were made with $\alpha=-0.9$ and with $\alpha$ free.  The results for all fits are given in tables~\ref{tab:schechter_fits} and \ref{tab:schechter_fits_bcg}.


\begin{table}
\caption{Best fitting parameters of a Schechter function for each cluster including BCGs.}

\begin{tabular}{llll}
\hline \hline
& $K^{*}$ (lower limit, upper limit) & $\phi^{*}$ & Prob($\chi^{2}$) \\ \hline
Cl0152 & 17.59 (17.26,17.90) & 47.89 & 0.56\\
Cl1226 & 17.79 (17.52,18.03) & 57.37 & $<10^{-4}$\\
Cl1415 & 17.96 (17.63,18.26) & 42.53 & 0.80\\
Combined ($\alpha_{{\rm fixed}}=-0.9$) & 17.81 (17.51,18.07) & 44.69 & $<10^{-4}$\\
Combined ($\alpha_{{\rm free}}=-0.94$) & 17.83 (17.1,18.3) & 45.34 & $<10^{-4}$\\
\hline \hline
\end{tabular}
\label{tab:schechter_fits}
\end{table}

\begin{table}
\caption{Best fitting parameters of a Schechter function for each cluster excluding BCGs.}

\begin{tabular}{llll}
\hline \hline
& $K^{*}$ (lower limit, upper limit) & $\phi^{*}$ & Prob($\chi^{2}$) \\ \hline
Cl0152 &17.76 (17.43,18.07) & 50.31 & 0.67\\
Cl1226 &18.04 (17.79,18.28) & 61.01 & 0.51\\
Cl1415 &18.08 (17.75,18.38) & 43.80 & 0.63\\
Combined ($\alpha_{{\rm fixed}}=-0.9$) &17.98 (17.68,18.26) & 45.59 & 0.38\\
Combined ($\alpha_{{\rm free}}=-0.54$) &18.53 (18.0,18.9)  & 79.66 & 0.52\\
\hline \hline
\end{tabular}
\label{tab:schechter_fits_bcg}

\end{table}

\begin{figure}
\begin{minipage}[c]{0.5\textwidth}
    \centering \includegraphics[scale=0.4,angle=270]
    {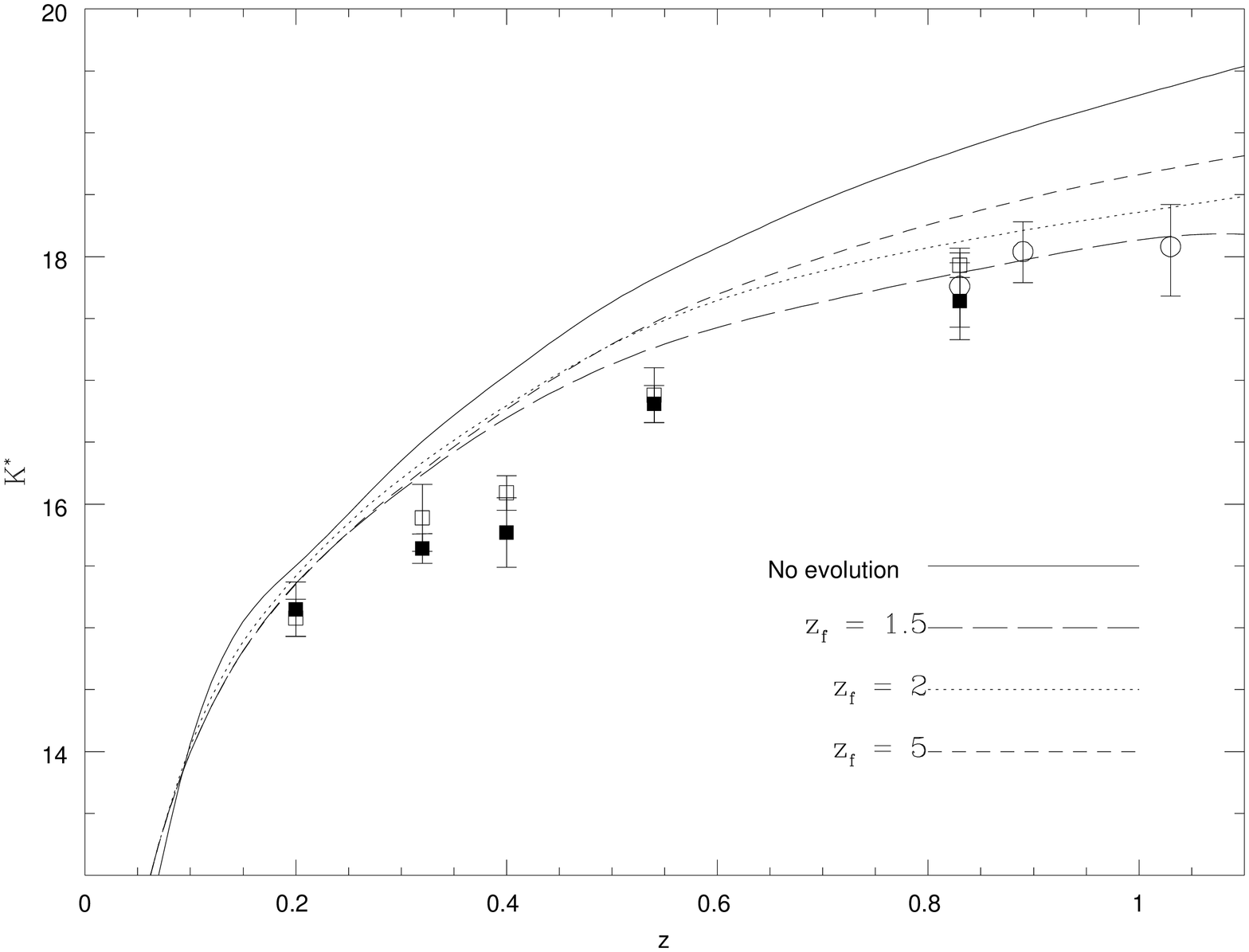}
  \end{minipage}
  \caption{The evolution of $K^{*}$.  The circles are data from this paper.  The squares are from de Propris et al. (1999), open symbols being low $L_{\mathrm{X}}$ systems and closed symbols being high $L_{\mathrm{X}}$ systems.}
  \label{fig:klf_evol}
\end{figure}

The evolution of $K^{*}$ is shown in figure~\ref{fig:klf_evol} along with the high $L_{\rm{X}}$ data from de Propris et al. (1999).  Models were computed for no-evolution, and passively evolving stellar populations with redshifts of formation $z_{\rm{f}}=1.5$, 2 and 5.  It can be seen that the bulk of the galaxies were significantly brighter in the past, than no-evolution predictions.  This is consistent with a passively evolving population with $z_{\rm{f}} \approx 1.5$--2.

\section{Conclusions}

The evolution of the galaxy populations of three high redshift clusters of galaxies has been studied.  The bulk evolution of the galaxies, as characterised by $K^{*}$, is found to be consistent with passive evolution with a redshift of formation $z_{\mathrm{f}}\sim\ 1.5$--2.  Further evidence for passive evolution is seen in the similarity of the shape of the high-redshift luminosity function with that of Coma.

Purely passive evolution of early-type galaxies is consistent with several other studies including the evolution of the $K$ band luminosity
function (de Propris et al. 1999), evolution of mass-light ratios (van Dokkum et al. 1998), and studies of
the scatter of the colour-magnitude relation (see eg.\ Ellis et al. 1997,
Stanford et al. 1998).

The lack of evolution observed in $K$ band luminosity function is in contrast to the conclusions of Kauffmann and Charlot (1998) and Kauffmann et al. (1996) for galaxies in the field. 

When discussing formation it is important to distinguish between the epoch at which the stars in the galaxies were formed and the epoch at which the galaxies were assembled.  If merging were a dissipationless process then it would be possible to have no extra star formation as a result of
a merger and thus the age of the stars within a galaxy could be older than the age of galaxy assembly.   A study of the cluster of galaxies MS 1054-03 at $z=0.83$ is presented by van Dokkum et al. (1999) in which there is observed a high fraction of merging red galaxies.  Very little star formation is seen in the merging galaxies constituting evidence that the galaxies are in fact somewhat younger than the stars that reside within them.  

Is such merging reflected in the evolution of the luminosity function?  The $K$
band luminosity of a galaxy is very nearly independent of star-formation, but reflects the mass of the old stars within the galaxy.  Thus $K$ magnitudes are a good measure of the stellar mass of a galaxy.  
 It is expected from semi-analytic models (Kauffmann et al. 1993, Baugh et al. 1996) that the typical redshift of assembly
of an elliptical will be higher in a rich cluster than in the field, due to the
fact that structures collapse earlier in denser environments.  This will be particularly relevant here since the systems investigated here are all very massive.  The semi-analytic models of Diaferio et al. (2001) predict very little evolution in the numbers of massive cluster galaxies since $z=0.8$ in a hierarchical model with dissipationless merging.  A high redshift of assembly of massive galaxies is in qualitative agreement with our results.
  We conclude that the luminosity evolution of bright galaxies in massive clusters is consistent with pure passive evolution, but note that this may be consistent with hierarchical models if most merging takes place at high redshifts. 

\end{document}